\documentclass[12pt]{article}
\begin{document}
\sf
\begin{center}
   \vskip 2em
  {\Large \sf Membrane geometry with auxiliary variables 
and quadratic constraints}
\vskip 3em
 {\large \sf  Jemal Guven \\[2em]}
\em{
Instituto de Ciencias Nucleares,
 Universidad Nacional Aut\'onoma de M\'exico\\
 Apdo. Postal 70-543, 04510 M\'exico, DF, MEXICO\\[1em]
and\\[1em]
School of Theoretical
Physics, Dublin Institute for Advanced Studies \\
 10 Burlington Road, Dublin 4, IRELAND \\
}
\end{center}
 \vskip 1em
\begin{abstract}
\sf Consider a surface described by a Hamiltonian which depends only
on the metric and extrinsic curvature induced on the surface.
The metric and the curvature, along
with the basis vectors which connect them to the embedding
functions defining the surface, are introduced as auxiliary
variables by adding appropriate constraints, all of them
quadratic. The response of the Hamiltonian 
to a deformation in each of the variables is examined
and the relationship between the multipliers implementing the
constraints and the conserved stress tensor of the theory established.
\end{abstract}
\vskip3em


Geometrical surfaces occur as representations of physical systems
across a spectacular range of scales spanning string theory,
cosmology, condensed matter and biophysics
\cite{SackLip,Lubensky,Vilenk,Polchi,Polyakov}. While the physics
they describe may be very different, the models involved share a
common feature: the action or Hamiltonian describing the surface
is constructed out of simple geometrical invariants of the surface
and fields which couple to it.
A nice example, close to home, is provided by a fluid membrane
consisting of amphiphilic molecules which aggregate spontaneously
into bilayers in water; at mesoscopic scales the membrane is
described surprisingly well by a Hamiltonian proportional to the
integrated square of the mean curvature \cite{CanHel,Reviews}. A 
close Lorenzian analog was introduced to describe color flux tubes in
QCD \cite{Polyakov,Kleinert}. There is now an extensive literature 
on the field theory of geometrical
models of this kind; a good point of
entry is provided by the review articles collected in
\cite{Nel.Pir.Wei:89,LesHouches}.

While the relevant geometrical model itself may be easy to
identify, typically it will involve derivatives higher than first
and inherit a level of non-linearity from the geometrical
invariants of the surface. There is, however, a useful strategem to 
lower the effective order 
or to tame this non-linearity involving the introduction of auxiliary
fields. In the description of a surface by a set of embedding
functions ${\bf X}$, the metric
induced on the surface is often replaced by an auxiliary 
intrinsic metric $g_{ab}$
\cite{David,Polyakov}; by amending the Hamiltonian with the
appropriate constraints, $g_{ab}$ is freed to be varied
independently of ${\bf X}$ \cite{Hamilt}.

Whereas the introduction of $g_{ab}$ as an auxiliary variable may
be sufficient for the technical purposes originally contemplated
--- providing a tractable inroad on the evaluation of a functional integral ---
from a purely geometrical point of view it is natural to question why
one should stop with the metric. In this paper, I will explore the
possibility of introducing additional auxiliary variables. Is it
possible, for example, to treat the extrinsic curvature as an
independent geometrical variable? This would be useful in
geometric theories involving higher derivatives.

Consider, for simplicity, a hypersurface with a single normal
vector. The extrinsic curvature $K_{ab}$ is defined in terms of
the behavior of this vector as it ranges over the surface;
together with the metric, it completely characterizes the surface
geometry. If $K_{ab}$ could be treated, like $g_{ab}$, as an
auxiliary field, the original theory describing a surface would
be replaced by a simple tensor field theory for $K_{ab}$ on a
curved space described by $g_{ab}$. The subtlety, of course, now lies
in the implementation of the constraints. The surprise is that it
is possible to do this in a way which is not only tractable but
also, en route, reveals a structure inherent to any theory of
embedded surfaces. Of course, if the constraints themselves were
to introduce new non-linearities the value of the exercise would be very
limited. This would certainly be the case in their implementation
within a functional integral \cite{Polyakov}. In this respect,
the metric tensor provides a useful set of auxiliary variables
because the induced metric depends quadratically on the first
derivatives of the ${\bf X}$. In contrast, as things stand, the
constraints involved in $K_{ab}$'s promotion to auxiliary
variable status are not quadratic.
Fortunately it is simple to resolve this difficulty: the basis 
vectors, the normal
and the tangents, are themselves introduced as intermediate 
auxiliary variable and the
constraint defining $K_{ab}$ is implemented (not in one but) in
a sequence of steps each of which involves a quadratic.
In a translationally invariant theory, ${\bf X}$ only appears though 
the tangent vectors; in such a theory, ${\bf X}$ is now 
consigned to the constraint defining the tangents and 
will appear nowhere else.

With the constraints in place it is possible to consider the
response of the Hamiltonian to deformations of each of these
variables in turn: for ${\bf X}$ the Euler-Lagrange derivative is
a divergence; in equilibrium this gives the conservation law
associated with translational invariance; the stress tensor gets
identified with the multipliers implementing the tangential
constraints. The auxiliary variables dispatch the task of
constructing this tensor in two clearly defined steps: first, the
Euler-Lagrange equations for the basis vectors  
express it in terms of the remaining multipliers;
these multipliers are then fixed by the Euler-Lagrange equations
for the metric and the curvature. The procedure is completely independent of
the details of the particular model. As described here, its
implementation depends in a unvarying way on each of the auxiliary
variables.

It is worthwhile to contrast the above picture with the more
familiar one which results when the metric alone is treated as an
auxiliary variable. The stress tensor coupling to the intrinsic
geometry is identified with the Lagrange multipliers implementing
the corresponding constraints. However, this tensor will {\it
not} generally be conserved: the induced metric characterizes
only the intrinsic geometry of the surface; there remains
considerable freedom as to how the surface is embedded in its
surroundings. The remaining multipliers capture this missing
information permitting the reconstruction of the full conserved
stress tensor underlying the geometry. The metric is just one
element in the complete description.

The geometry of interest is a $D$-dimensional surface embedded in
$R^{D+1}$ described by ${\bf x} = {\bf X} (\xi^1,\dots,\xi^D)$.
Higher co-dimensions will not be considered though it is
straightforward to do so; it is also straightforward to adapt the
discussion to consider time-like surfaces in Minkowski space.
Indeed, the description may also be extended to surfaces in a
curved background. The notation used is ${\bf
x}=(x^1,\dots,x^{D+1})$; the parameters $\xi^1,\dots,\xi^D$
represent local coordinates on the surface. One now shifts the
focus of attention from the embedding functions ${\bf X}$ to the
geometrical tensors induced by them, the metric and the extrinsic
curvature (for example, see \cite{Spivak})
\begin{equation}
g_{ab}= {\bf e}_a\cdot {\bf e}_{b}\,; \quad K_{ab}= {\bf
e}_a\cdot \partial_b{\bf n}\,, \label{eq:gKdef}
\end{equation}
$a,b=1,\dots, D$, where ${\bf e}_a$ are tangent and ${\bf n}$ is
unit normal to the surface:
\begin{equation}
{\bf e}_a= \partial_a {\bf X}\,;\quad {\bf e}_a\cdot {\bf n}
=0\,;\quad  {\bf n}^2 =1\,. \label{eq:endef}
\end{equation}
Together, $g_{ab}$ and $K_{ab}$ encode the geometrically
significant derivatives of ${\bf X}$; all geometrical invariants,
the Hamiltonian included, can be cast as functionals of $g_{ab}$
and $K_{ab}$.

Consider any reparametrization invariant functional of the
variables $g_{ab}$ and $K_{ab}$,
\begin{equation}
H [{\bf X}]= \int dA \, {\cal H}(g_{ab}, K_{ab})\,. \label{eq:HX}
\end{equation}
The area element is $dA = \sqrt{ {\rm det} g_{ab}} \,d^D\xi$. We
are interested in determining the response of $H$ to a
deformation of the surface: ${\bf X}\to {\bf X}+ \delta {\bf X}$.
The approach adopted here will be to distribute  the  burden on
${\bf X}$ among ${\bf e}_a,{\bf n},g_{ab}$ and $K_{ab}$ treating
the latter as independent auxiliary variables. To do this
consistently the structural relationships connecting the
variables must be preserved under the deformation; thus
Eqs.(\ref{eq:gKdef}) defining $g_{ab}$ and $K_{ab}$ in terms of
the basis vectors ${\bf e}_a$, and ${\bf n}$, as well as
Eqs.(\ref{eq:endef}) which define these vectors are introduced as
constraints; $H$ is amended accordingly.

Introduce Lagrange multiplier functions  to implement the
constraints. We thus construct a new functional $H_C [g_{ab},
K_{ab}, {\bf n}, {\bf e}_a, {\bf X},{\bf f}^a,\Lambda^{ab},
\lambda_{ab},\lambda_\perp^a,\lambda_n]$ as follows:
\begin{eqnarray}
H_C&=& H [g_{ab}, K_{ab}]+   \int dA \,\, {\bf f}^a \cdot ({\bf
e}_a-\partial_a{\bf X}) \nonumber\\
&& \quad \quad\quad + \int dA\,\, \left(\lambda_\perp^a ({\bf
e}_a\cdot {\bf n}) + \lambda_n ({\bf n}^2 -1) \right)
\label{eq:Fdef}
\\
&&+ \int dA \,\,\left(\Lambda^{ab} (K_{ab}- {\bf e}_a\cdot
\partial_b{\bf n})
+ \lambda^{ab}(g_{ab}- {\bf e}_a\cdot{\bf e}_b)\right) \,.
\nonumber
\end{eqnarray}
Note that the original Hamiltonian $H$ is now treated as a
function of the independent variables, $g_{ab}$ and $K_{ab}$ but
not of ${\bf e}_a,{\bf n}$ or ${\bf X}$. The multiplier ${\bf
f}^a$ anchors ${\bf e}_a$ to the embedding ${\bf X}$; it is
simultaneously a spatial vector and a surface vector. Its
geometrical character is dictated by the constraint it imposes.
Likewise, the multipliers $\Lambda^{ab}$ and $\lambda^{ab}$ are
symmetric surface tensors; $\lambda_\perp^a$ is a surface vector,
and $\lambda_n$ is a scalar. We are now free to treat $g_{ab},
K_{ab}$, ${\bf n}$, ${\bf e}_a$ and ${\bf X}$ as independent
variables which can be deformed independently. It is not
necessary to track explicitly the deformation induced on $g_{ab}$
and $K_{ab}$ by a deformation in ${\bf X}$.

The only place where ${\bf X}$ appears explicitly in $H_C$ is
within the constraint which defines ${\bf e}_a$. The
corresponding Euler-Lagrange derivative is a divergence
\begin{equation}
\delta H_C/\delta {\bf X}
=\nabla_a {\bf f}^a\,. \label{eq:nabf}
\end{equation}
In this expression $\nabla_a$ is the symmetric covariant
derivative compatible with $g_{ab}$ and operates on surface
indices. In equilibrium, ${\bf f}^a$ is covariantly conserved on
the surface. The physical interpretation of ${\bf f}^a$ as a
stress tensor will be commented on below.

The Euler-Lagrange equations for ${\bf e}_a$ express the
conserved `vector' ${\bf f}^a$ as a linear combination of the
basis vectors:
\begin{equation}
{\bf f}^a = (\Lambda^{ac}\, K_{c}{}^b   +  2\lambda^{ab}) {\bf
e}_b - \lambda_\perp^a {\bf n}\,. \label{eq:e2}
\end{equation}
The Weingarten equations $\partial_a{\bf n}= K_{a}{}^b {\bf e}_b$
have been used to obtain Eq.(\ref{eq:e2}).
They themselves follow from the constraints on $K_{ab}$ and the
normalization of ${\bf n}$. 

Remarkably, ${\bf f}^a$ is determined in a model independent way
in terms of the Lagrange multipliers imposing the geometrical
constraints.
The values assumed by the multipliers will, of course, depend on
the specific Hamiltonian $H$.

The multiplier $\lambda^a_\perp$ enforcing orthogonality
appearing in Eq.(\ref{eq:e2}) is fixed by the Euler-Lagrange
equation for ${\bf n}$.
Using the Gauss equations $\nabla_a {\bf e}_b =- K_{ab} {\bf n}$
(which themselves follow from the Weingarten equations and the
orthogonality constraint),
one has
\begin{equation}
(\nabla_b \Lambda^{ab} + \lambda_\perp ^a) {\bf e}_a + (2
\lambda_n  - \Lambda^{ab} K_{ab}) {\bf n} =  0\,, \label{eq:n}
\end{equation}
and thus
\begin{eqnarray}
\lambda_\perp^a &=& -\nabla_b \Lambda^{ab}\;\\
2\lambda_n  &=& \Lambda^{ab} K_{ab}\,. \label{eq:lamn}
\end{eqnarray}
$\lambda_\perp^a$ is identified as (minus) the divergence of
$\Lambda^{ab}$; the normal component of ${\bf f}^a$ will generally
involve one derivative more than its tangential components. Note
that $\lambda_n$ does not appear in the stress tensor. This is
not surprising: the role of $\lambda_n$ is to enforce the
normalization of ${\bf n}$, which is important for reasons of
mathematical consistency but not physically.

The  missing ingredients are the multipliers $\Lambda^{ab}$ and
$\lambda^{ab}$ appearing in the tangential part of ${\bf f}^a$.
They are determined by the Euler-Lagrange equations for $K_{ab}$
and $g_{ab}$:
\begin{eqnarray}
\Lambda^{ab} &=& -{\cal H} ^{ab}\\
\lambda^{ab} &=&  T^{ab}/2\,,
\end{eqnarray}
where ${\cal H}^{ab} = \partial {\cal H}/\partial K_{ab}$ and
$T^{ab} =- 2(\sqrt{g})^{-1} \partial (\sqrt{g}{\cal H})/ \partial
g_{ab}$ is the intrinsic stress tensor associated with the metric
$g_{ab}$. The conserved stress ${\bf f}^a$ is
\begin{equation}
{\bf f}^a = (T^{ab} - {\cal H}^{ac}\, K_{c}{}^b)\, {\bf e}_b -
\nabla_b {\cal H}^{ab} \,{\bf n}\,.
\end{equation}
Note that $T^{ab}$ is only one part of the total stress tensor and
it is entirely tangential; it is not generally conserved.

There is no difficulty treating a Hamiltonian of the more general
form ${\cal H}(g_{ab}, K_{ab},\nabla_a K_{bc}, \dots)$ within
this framework; the derivatives appearing in $T^{ab}$ and ${\cal
H}^{ab}$ are simply replaced by functional derivatives. It is
also unnecessary to consider an explicit intrinsic curvature
dependence in ${\cal H}$. This is because the Gauss-Codazzi
equations \cite{Spivak}
\begin{equation}
R_{abcd}= K_{ac} K_{bd} - K_{ad} K_{bc} \label{eq:GC}
\end{equation}
completely fix the Riemann tensor in terms of the extrinsic
curvature.

Now let us look at a few examples. For a soap film, or a
Dirac-Nambu-Goto membrane, $H$ is proportional to the surface
area with a constant surface tension $\mu$: ${\cal H}^{ab}=0$,
and $T^{ab}= -\mu g^{ab}$; the stress is determined completely by
the metric; the only relevant constraints are intrinsic. A less
simple example is provided by the Helfrich Hamiltonian without
adornment describing a fluid membrane with ${\cal H}= \alpha K^2
+ \mu$ in Eq.(\ref{eq:HX}), where $K = g^{ab} K_{ab}$. The first
term, a conformal invariant when $D=2$, was introduced by
Willmore\cite{Willmore}.
One has ${\cal H}^{ab}= 2 \alpha g^{ab} K$, and $T^{ab}= \alpha K
(4  K^{ab}-  K g^{ab}) - \mu g^{ab} $. Thus
\begin{equation}
{\bf f}^a = [\alpha K(2 K^{ab}  - K g^{ab} )- \mu g^{ab}]\, {\bf
e}_b - 2\alpha \nabla^a K \, {\bf n}\,.
\end{equation}
In general, if ${\cal H}$ does not involve derivatives of
$K_{ab}$, as is the case in the description of a fluid membrane,
neither will $\Lambda^{ab}$ or $\lambda^{ab}$. Thus the
tangential component of ${\bf f}^a$ will not involve derivatives
of curvatures.

Eq.(\ref{eq:nabf}) casts the Euler-Lagrange equations for ${\bf
X}$ as a conservation law, $\nabla_a {\bf f}^a=0$. Following
Refs.\cite{Stress}, write
\begin{equation}
{\bf f}^a = f^{ab} {\bf e}_b + f^a {\bf n}\,.
\end{equation}
The projections of Eq.(\ref{eq:nabf}) normal and tangent to the
surface give respectively:
\begin{eqnarray}
\nabla_a f^a - K^{ab} f_{ab} &=&0\,,\label{eq:nor}\\
\nabla_a f^{ab} + K^{ab} f_a &=& 0\,.\label{eq:tan}
\end{eqnarray}
Eq.(\ref{eq:nor}) is the `shape' equation. For the example
considered above, it reads \cite{Hel.OuY:87}
\begin{equation}
- 2\alpha\nabla^2 K - \alpha K K^{ab} ( 2 K_{ab} - K g_{ab}) +
\mu K =0\,. \label{eq:shape}
\end{equation}
Because $H$ is invariant under reparametrizations, the only
physical deformations are those normal to the surface. There is a
single `shape' equation \cite{Stress}. Eqs.(\ref{eq:tan}) are
consistency conditions on the components of the stress tensor.
For a Hamiltonian invariant under reparametrizations, they reduce
to trivial geometrical identities.

This framework also provides a physical interpretation of the
conserved multiplier ${\bf f}^a$. Look at the divergence that was
legitimately discarded in the derivation of the Euler-Lagrange
equations: modulo these equations, the deformed Hamiltonian is
\begin{equation}
\delta  H_C =   -\int dA \, \nabla_b (\Lambda^{ab} {\bf e}_a\cdot
\delta {\bf n} + {\bf f}^b \cdot \delta {\bf X})\,.
\label{eq:delHN}
\end{equation}
A spatial translation $\delta {\bf x} = {\bf a}$, where ${\bf a}$
is some constant vector, induces the internal symmetry  $\delta
{\bf X}= {\bf a}$; all of the other variables are unchanged. In
particular, $\delta {\bf n}=0$ in Eq.(\ref{eq:delHN}). Thus
\begin{equation}
\delta  H_C =   -{\bf a} \cdot \int dA \, \nabla_a {\bf f}^a \,.
\end{equation}
On a domain $\Sigma$ with boundary the left hand side
may be cast as an integral over this boundary;
the vector $\eta_a {\bf f}^a dS$, where $\eta_a$ is the outward
normal to the boundary $\partial\Sigma$, is thus identified as
the force on the boundary element $dS$ due to the action of the
stresses ${\bf f}^a$ set up within the domain.

The construction of the stress tensor for a fluid membrane was
considered some time ago by Evans in a bio-mechanical context
\cite{Eva:74}. In \cite{Stress}, the problem was reconsidered
from a geometrical perspective, and the stress tensor identified
as the conserved Noether current associated with translational
invariance. This was done by tracking the response of the metric
and extrinsic curvature to the deformation  in the embedding
functions. The approach via auxiliary variables, adopted here,
has the virtue of sidestepping the need to know how $g_{ab}$ or
$K_{ab}$ themselves respond to a deformation in ${\bf X}$ and the
attendant problem of doing so in a way which respects the
invariance under change of parametrization.

A few technical comments on the choice of constraints:

\noindent (1) All of the constraints are bi-linear in
the vectors ${\bf e}_a$ and ${\bf n}$ with one exception --- the
linear constraint, ${\bf e}_a=
\partial_a{\bf X}$. It would be consistent to implement the
linear Gauss-Weingarten equations as vector constraints in place
of the bilinear definition of $K_{ab}$. There is, however, a sound
reason not to: with the bi-linear choice of constraint used here,
${\bf f}^a$ gets identified directly as a linear combination of
${\bf e}_a$ and ${\bf n}$.

\noindent (2) It is consistent to use a reduced set of auxiliary
variables; for example, the tangent vectors ${\bf e}_a$ or the
curvatures $K_{ab}$ could be dropped. In the former case, instead
of implementing ${\bf e}_a= \partial_a{\bf X}$ as a constraint,
substitute in favor of ${\bf X}$ everywhere ${\bf e}_a$ appears.
If one also drops $K_{ab}$ as an independent variable, then $K^2
= (\nabla^2 {\bf X})^2$ \cite{David}; for the Helfrich
Hamiltonian, ${\bf n}$ does not appear so it is also consistent
to drop the constraints involving ${\bf n}$. The only remaining
auxiliary variable is the metric: the original auxiliary variable
inspiring this generalization. The disadvantage of this
truncation is that the constraint ${\bf e}_a= \partial_a {\bf X}$
comes with the marker identifying the stress tensor and, when it
is dropped, with it goes the conservation law encoded in
(\ref{eq:nabf}) as well as the identification of the conserved
stress tensor appearing within it.

\noindent (3) If one attempts to treat $g_{ab},
K_{ab},{\bf n}, {\bf e}_a$ and  ${\bf X}$ as independent
variables with an insufficient set of constraints, an
inconsistent set of equations is usually obtained. For example,
had the normalization constraint been dropped, instead of
identifying $\lambda_n$, Eq.(\ref{eq:lamn}) would have given
$\Lambda^{ab} K_{ab}=0$ which is nonsense unless, of course,
$\Lambda^{ab}$ itself turns out to be zero --- as it does for the
soap film.

\noindent (4) There is no need to implement the
Gauss-Codazzi, or the Codazzi-Mainardi integrability conditions
explicitly as constraints. Recall that the former are given by
Eqs.(\ref{eq:GC}); the latter are $\nabla_{a} K_{bc}- \nabla_b
K_{ac}=0$. When the constraints appearing in Eq.(\ref{eq:Fdef})
are satisfied, the integrability conditions are automatically
accounted for. One might choose to focus, however, on a specific
parametrization of $g_{ab}$ or $K_{ab}$ (asymptotic coordinates,
for example) which is not anchored to a specific embedding. In
such a case, consistency would require the implementation of the
integrability conditions as additional constraints.

To conclude, a geometrical framework involving auxiliary
variables has been introduced to examine a theory of surfaces
described by a reparametrization invariant Hamiltonian,
exemplified by the Helfrich Hamiltonian describing fluid
membranes. For this Hamiltonian, the only variable which appears
in a non-quadratic way in $H_C$ is the metric. This would suggest
that the approach has the potential to provide novel
approximations to geometrical functional integrals. Because of
the central role played by the stress tensor, it should also
prove useful in the study of membrane mediated interactions
\cite{protein}; this would certainly appear to be the case if
non-perturbative effects are important and it becomes necessary
to look beyond the quadratic truncation of the Hamiltonian in
terms of the height function. By implementing geometrical
constraints using Lagrange multipliers, it is possible to
establish useful connections between models for embedded surfaces
and other, more fully studied or more tractable, models. For
example, it is possible to consider the Helfrich model as a
constrained $O(D+1)$ non-linear sigma model on the surface
\cite{Polyakov}. The Hamiltonian density is $(\nabla_a {\bf
n})^2$ subject to the constraint ${\bf e}_a\cdot {\bf n}=0$ on
the unit vector. Applications, as well as generalizations, will
be considered in forthcoming publications.

\vskip1em\noindent
{\sf Acknowledgements} 

\vskip1em\noindent
I have benefited from discussions with Riccardo
Capovilla, Chryssomalis Chyssomalakos and Denjoe O' Connor. I also
thank DO'C
for his hospitality during my stay at DIAS. Partial support from
DGAPA-PAPIIT grant IN114302 is acknowledged.

\end{document}